\documentclass[twocolumn, article, floatfix, a4paper, 12pt]{article}

\usepackage{graphicx}
\usepackage{amsmath,amssymb,amsfonts}
\usepackage{color}
\usepackage{hyperref}
\usepackage{textcomp}
\usepackage{times}
\usepackage{epstopdf}
\usepackage{color}
\newcommand{\Keywords}[1]{\par\noindent
{\small{\em Keywords\/}: #1}}

\title{Imaging material properties of biological samples with a Force Feedback Microscope}
\author{\normalsize{Luca Costa$^{1,2}$, Mario S Rodrigues$^3$, Emily Newman$^4$, Chloe Zubieta$^1$, Jo\"{e}l Chevrier$^{1,2,5}$, Fabio Comin$^1$} \\
\begin{minipage}{.80 \linewidth}
\small
$^{1)}$ \textit{European Synchrotron Radiation Facility, 6 rue Jules Horowitz BP 220, 38043 Grenoble Cedex, France} \\
$^{2)}$ \textit{Universit\'{e} Joseph Fourier BP 53, 38041 Grenoble Cedex 9, France} \\
$^{3)}$ \textit{CFMC/Dep. Fisica, Faculdade de Ci\^encias, Universidade de Lisboa, Campo Grande, 1749-016 Lisboa, Portugal} 	\\
$^{4)}$ \textit{European Molecular Biology Laboratory, 6 rue Jules Horowitz BP 181, 38041 Grenoble Cedex 9, France} 	\\
$^{5)}$ \textit{Institut N\'{e}el CNRS BP 166, 38042 Grenoble Cedex 9, France} \\ 
\end{minipage} \\
\begin{minipage}{.85 \linewidth}
\small
Mechanical properties of biological samples have been imaged with a \textit{Force Feedback Microscope}.
Force, force gradient and dissipation are measured simultaneously and quantitatively, merely knowing the AFM cantilever spring constant. 
Our first results demonstrate that this robust method provides quantitative high resolution force measurements of the interaction
The little oscillation imposed to the cantilever and the small value of its stiffness result in a vibrational energy much smaller than the thermal energy, reducing the interaction with the sample to a minimum. We show that the observed mechanical properties of the sample depend on the force applied by the tip and consequently on the sample indentation.
 Moreover, the frequency of the excitation imposed to the cantilever can be chosen arbitrarily, opening the way to frequency-dependent studies in biomechanics, sort of spectroscopic AFM investigations.\\
\end{minipage}
}
\date{\small{\today}}

\begin{document}
\maketitle
\section{Introduction}
Since years Atomic Force Microscopy (AFM) is a powerful technique for analyzing the physical properties of materials down to nanoscale.
More recently AFM has made its entry in the biological world, where the fragile nature of samples has prompted new challenges and has led to considerable optimization of the AFM techniques\cite{raman,tando,tetard}.
One issue is linked to the softness of the samples with respect to the AFM tip that imposed the minimization of the contact between tip and sample and a second one is the increasingly urgent need of extracting quantitative values for the mechanical and chemical properties of the studied systems. In this letter we address specifically these two aspects.

In conventional atomic force microscopy a cantilever with a 
nanosized tip is used to explore the entire range of tip-sample forces in one single vibrational cycle. 
The tip-sample interaction does couple the eigenmodes of the cantilever but typically only the information contained in the first eigenmode is studied.
However, the coupling with higher modes is highly nonlinear for relatively large amplitudes of oscillation and energy is transferred to higher harmonics \cite{stark:5111, stark347}. 
Consequently, if the cantilever response is measured only at the excitation frequency, close to the first eigenmode, then part of the tip-sample interaction is masked and not measured. 
To overcome the problem, methods have been developed where several modes and/or harmonics are measured contemporaneously \cite{raman, kareem, garcia}. 
Now, among the difficulties in dealing with biological sample is the fact that the measurements are usually carried out in liquid.
When imaging in liquid the AFM cantilevers have to be stiff enough to maintain an acceptable quality factor to run dynamic AFM measurements. This, together with the large amplitudes of oscillation imposed, results in large excitation energies when compared to thermal energy. 
Exciting also at other frequencies further increases the excitation energy and the consequent energy transfer to the sample via the tip-sample interaction.
For decreasing both the excitation energy and the pressure exerted on the sample by the tip, it becomes of paramount importance then to decrease both the cantilever stiffness and the amplitude of oscillation. Moreover, small oscillation amplitudes result in a negligible coupling to higher harmonics. 

Recently this strategy has been implemented in a new instrument called \textit{Force Feedback Microscope}(FFM) \cite{rodrigues:203105} where very soft cantilevers and small amplitudes of oscillation are adopted to minimize the interaction energy.
The cantilever stiffness is typically kept in the order of $0.01$ N/m and the oscillation amplitude is about $0.3$ nm.  
The typical excitation energy imposed to the cantilever is then
$E=k x^2 \approx 10 \times 10^{-22}J$
while $k_b T \approx 41 \times 10^{-22}$, implying that the excitation energy is kept below the thermal energy.
TThis can be compared to normal AFM measurements where the amplitude and the stiffness are at least a factor 10 higher ($3$ nm and $0.2$ N/m) which make a factor 1000 in energy.
Moreover, using small amplitudes of oscillation also offers the advantage that it can be assumed that at any given distance 
the tip-sample interaction is linear justifying the use of very simple equations to describe the interaction. In turn this has the consequence that the changes in normalized oscillation amplitude and phase can be mapped directly into stiffness and damping of the sample.
One central aspect of FFM is that rather than 
\emph{assuming} a certain dynamic behavior of the cantilever, 
it is possible to \emph{calibrate} its dynamics as a function of a measured reference interaction. 
This makes it possible to quantify easily the tip-sample interaction regardless of the cantilever response spectrum. In liquid conditions and in particular when soft cantilevers are used, it is often difficult to precisely obtain a quality factor (Q) or even identify the resonance frequency $f$ of the cantilever. These two constants are essentially irrelevant when using the method described here. The frequency used during the measurements is arbitrarily chosen and kept constant during a measurement. The responses in frequency of the liquid and of other mechanical parts do not influence the quantitative analysis.

The results reported on this letter show how the FFM makes possible to map the topography, the force, the force gradient and the dissipation in one single scan, and how the interaction can be measured quantitatively from solely the knowledge of the spring constant of the cantilever.
The range of the xyz scanner used was rather large ($100\times 100 \times 100 \mu m$) limiting the spatial resolution.
This does not limits the significance of our results, since our main goal is to demonstrate the possibilities offered by the method.
We used three different samples: DNA, lipids and protein complexes in liquid media. For all the three samples the substrate was mica.  
\section{Materials and method}
\subsection{Force Feedback Microscopy}
\begin{figure}[htp]
 \centering
  \includegraphics[width=\linewidth]{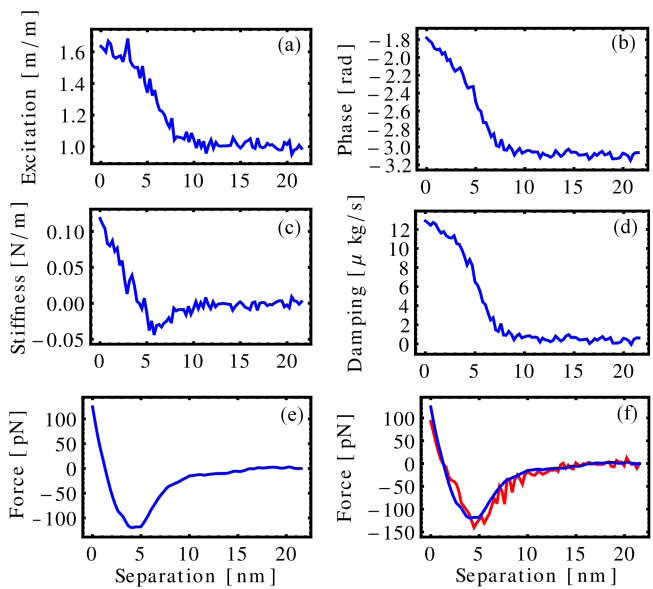} %
 \caption{Approach curve for calibration of the force sensor. 
 The sample is clusters of TBK1 and OPTN complexes on mica.
 (a) normalized excitation (b) phase difference, 
 (c) tip-sample stiffness (d) damping coefficient,
 (e) negative of integrated force gradient and (f) force (red) 
 comparison with tip-sample stiffness (blue, thinner line).
 }
 \label{fig:1}
 \end{figure}

Before taking an image, a set of approach curves onto the mica substrate are performed to calibrate the cantilever dynamics.
A typical curve is shown in figure \ref{fig:1}.
The FFM feedback loop keeps the position of the tip constant relative to the laboratory reference frame. 
The force supplied by the loop is then equal and opposite to the tip-sample interactions \cite{rodrigues:203105}.
The calibration is a measurement of how the cantilever responds elastically and inelastically to forces at the frequency and in the medium chosen.
To perform the calibration the oscillation amplitude, the excitation amplitude and the phase are recorded as a function of the distance (or interaction) resulting in a so called approach curve. 
\begin{equation}
 \nabla F = a \left[\cos(\phi_\infty)-n \cos(\phi)\right]
 \label{eq:eq1}
\end{equation}
\begin{equation}
 \gamma = \frac{a}{\omega}\left[\sin(\phi_\infty)-n \sin(\phi)\right]
 \label{eq:eq2}
\end{equation}
Equations \ref{eq:eq1} and \ref{eq:eq2} are used to convert measured data to interaction \cite{rodrigues:203105} parameters namely to force gradient $\nabla F$ and viscous damping $\gamma$.
The tip-sample force gradient corresponds to the negative of the tip-sample stiffness and for that reason we may use stiffness or force gradient to refer to the same physical characteristics of the interaction.
In the equations above $a$ and $\phi$ are calibrated constants,
$n$ is the normalized amplitude (i.e. the ratio excitation amplitude to oscillation amplitude normalized to one at infinity) 
and $\omega$ is the angular velocity of the excitation.
The strategy consists in finding which constants, $a$ and $\phi_{\infty}$, satisfy the condition that the integral of the force gradient equals the force.
Since the force is simply $F=k \Delta x$, the only constant that is required for calibrating the cantilever dynamics at that specific frequency and in the specific media is the cantilever stiffness $k$
that allows to obtain the force.
To note that the force is actually the amount of force that the feedback loop needs to supply to the tip to maintain it at equilibrium.
The method to calibrate the cantilever is described in more detail in reference \cite{rodrigues:203105}.
To obtain the images a second feedback loop is used.
This second feedback loop operates in the same way as in any other typical AFM measurement, moving the sample to and fro maintaining constant a chosen signal, typically the amplitude of oscillation.
Here, instead of the amplitude of oscillation we have used either the phase of oscillation or the tip-sample force.
\subsection{DNA}
The sample was prepared using a solution containing Mg$_2$+ divalent cations to bind the DNA on top of freshly cleaved mica \cite{dna2,dna3,dna4}.
In our experiments we used 1000 base-pair DNA and supercoiled DNA.
A buffer $10$ mM HEPES, $5$ mM MgCl$_2$ at pH $6.5$ was used to dilute the DNA up to $1$ nM concentration.
A drop of  $10$ $\mu$L of the DNA solution has been deposited on freshly cleaved mica and left incubate for 20 minutes.
The drop is then rinsed with $500$ $\mu$L of $10$ mM HEPES, $5$ mM MgCl$_2$ at pH $6.5$ and the sample is imaged in buffer with the FFM.
\subsection{Phopholipids}
The phospholipid 1,2-Distearoyl-sn-glycero-3-phosphoethanolamine (DSPE) was used to obtain self-assembled lipids layers on mica.
Lipids were diluted in chloroform at a concentration of $0.1$ g/L.
About $20$ $\mu$L of solution was directly applied 
on freshly cleaved mica at room temperature.
The specimen was incubated for 15 minutes and then washed several times with deionized water.
The sample is imaged in $20$ mM Tris, $150$ mM NaCl at pH $7.5$ with the FFM.
\subsection{Tank Binding Kinase (TBK1) and Optineurin (OPTN) protein complexes}
The TBK1·OPTN sample was prepared by adding the two purified 
proteins in equimolar ratios and purifying the 1:1 complex via size exclusion chromatography. The complex was diluted with deposition buffer ($20$ mM HEPES and $5$ mM MgCl$_2$) to $34$ nM. Sample grids were prepared by applying $20$ $\mu$L of poly-L-lysine to 
freshly cleaved mica to render the surface positively charged \cite{proteins2, lipids5}.
After 5 min incubation the mica was rinsed with dH$_2$O and dried with gaseous nitrogen. 
Subsequently $2$ $\mu$L of the TBK1·OPTN sample was added to the mica and incubated for ten min. 
The mica was rinsed with deposition buffer and imaged with the FFM in non-contact mode.
\section{Results}
The first case we present here is that of DNA on mica.
Imaging DNA in a $\mathrm{MgCl_2}$ solution is particularly difficult. 
\begin{figure}[htp]
 \centering
  \includegraphics{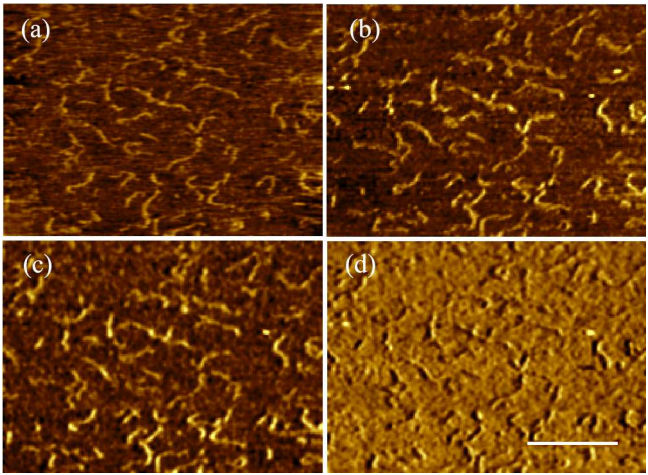} 
 \caption{FFM images of DNA deposited on mica in liquid solution.
 (a) topography, (b) force, (c) stiffness and (d) damping.
 The full color scale is 3 nm, 600 pN, 0.025 N/m and 1 $\mu$kg/s 
 respectively and the scale bar is 500 nm.
 Here the feedback signal used for imaging is the phase of oscillation in 
 repulsive regime yielding an almost constant damping image.}
 \label{fig:2}
 \end{figure}
\noindent 
Figure \ref{fig:2} shows the topography, force, stiffness and damping
images of DNA.
A cantilever with 0.02 N/m stiffness was used.
The excitation frequency was $3.555$ kHz and the oscillation amplitude of about $0.3$ nm.
To obtain the topography the phase difference between excitation and oscillation was used as set point in the topography feedback loop.
This corresponded to an image at constant damping.
Figure \ref{fig:2} shows the damping image resembling to an error image. The same figure also provides local force changes (Fig. 2b) and local stiffness changes (Fig. 2c).
We observe the interaction force to be close to zero when the tip is on the top of the mica. When the tip is on the top of DNA we observe an interaction force equal to $150$ pN. We conclude that the indentation on the DNA is therefore larger than the one on mica, indicating the DNA to be less viscous than mica. The measured local stiffness of DNA is larger than the one of the mica, indicating the DNA to be stiffer than mica \textbf{at this dissipation}.\\
\begin{figure}[htp]
 \centering
  \includegraphics[width=\linewidth]{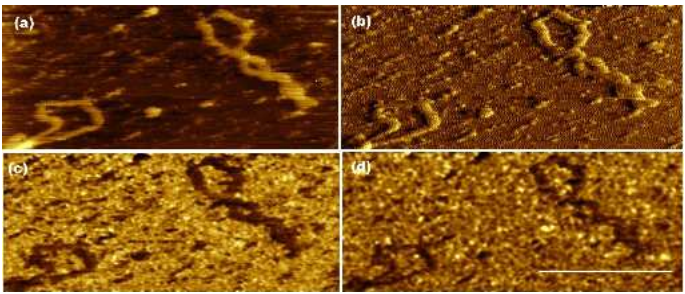} %
 \caption{FFM images of supercoiled DNA deposited on mica in liquid solution.
 (a) topography, (b) force, (c) stiffness and (d) damping.
 The full color scale is 2 nm, 300 pN, 0.13 N/m and 4.8 $\mu$kg/s 
 respectively and the scale bar is 400 nm.
 Here the feedback signal used for imaging is the force in 
 repulsive regime.
 }
 \label{fig:new}
 \end{figure}
In figure \ref{fig:new} supercoiled DNA has been imaged at a constant repulsive force of 100 pN. The excitation frequency was $3.57$ kHz, the oscillation amplitude was $0.3nm$ and the cantilever stiffness $0.02$ N/m. In this imaging mode we acquired the stiffness and the damping coefficient simultaneously to the topography. In this case the DNA is softer than mica of one order of magnitude (figure 3c). Moreover, DNA is less viscous than mica (figure 3d), in agreement with the measurement presented in figure \ref{fig:2}.
We conclude that the constant dissipation imaging mode can be seen as a measurement of the local stiffness for different interaction forces, since the damping coefficient is largely changing as a function of the interacting sample. Concluding, the local stiffness of the 1000 base-pair DNA at $150$ pN it is found to be $0.025$ N/m (figure 2), whereas for the supercoiled DNA at $100$ pN it is found to be $0.01$ N/m (figure 3).\\
As a second case we show an image of lipid membranes.
Phospholipids are the major components of all cell membranes, 
constituting the matrix for the membrane proteins.
Cells can perform many physiological functions through the membrane,  
including molecular recognition, intracellular communication and cell adhesion \cite{lipids1}, but the direct observation of these biological events at the nanoscale is still a challenge.
Simplified 2-dimensional systems called artificial membranes are used to simulate cell membranes.
These membranes assembled with phospholipids are intensively studied as a model for the cell membrane \cite{lipids3} and membrane/proteins interaction \cite{lipids4} with the AFM.
The measurements were performed at a constant repulsive force of $50$ pN.
A small oscillation amplitude of $0.2$ nm at $7.01$ kHz was imposed on the tip.
In the topography, figure 4a, the thickness of the DSPE layers is found to be 6.5 nm $\pm$ 1 nm, indicating the DSPE to form a bilayer \cite{lipidsa}.
The images clearly provide more information than just the topography.
In figure 4c for example the color contrast indicates when the tip is over the membranes as they appear locally softer than the substrate.
This is in agreement with the measurements performed in the last decade with the acquisition of static force curves and \textit{Peak Force} techniques \cite{lipidsb,lipidsc}. Moreover, we observe that thicker layers of lipids result to be softer and less viscous than a single bilayer. This is likely due to the lower influence of the substrate.
 \begin{figure}[htp]
 \centering
 \includegraphics[width=\linewidth]{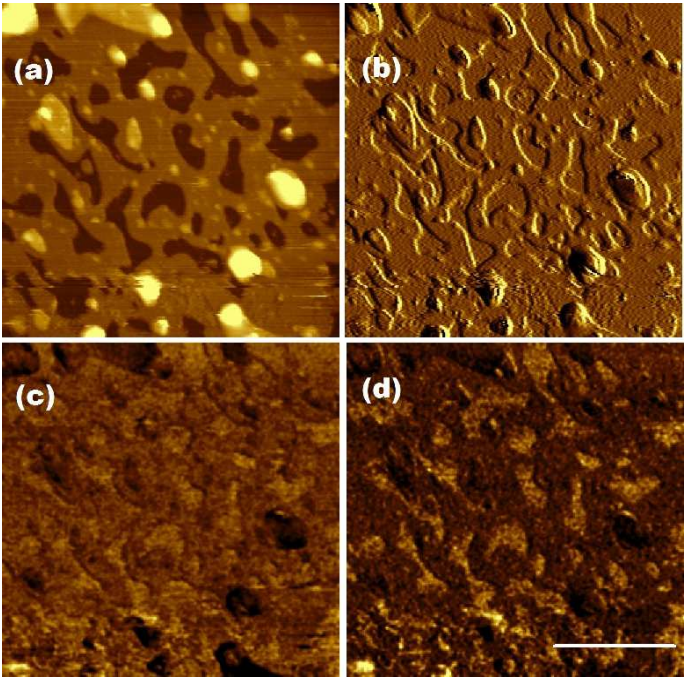} 
 \caption{FFM images of DSPE deposited on mica in liquid solution.
 (a) topography, (b) force, (c) stiffness and (d) damping.
 the full color scale is 23nm, 100pN, 0.13 N/m and 2.5$\mu$kg/s respectively. 
 The scale bar corresponds to 1000 nm.
 The signal chosen for the feedback was a small repulsive force of 50 pN.}
 \label{fig:3}
 \end{figure}
The last example we show is a non-contact image of clusters 
of the Tank Binding Kinase (TBK1) and Optineurin (OPTN) protein complexes. 
The characterization of biologically relevant protein-protein complex is essential for understanding fundamental cellular processes. 
The TBK1 is a vital protein involved in the innate immune 
signaling pathway. TBK1 forms a complex with the scaffold protein OPTN. This complex (TBK1·OPTN) has not yet been characterized structurally due to its large size and intrinsic flexibility. 
Structural characterization would help to elucidate how the 
complex is involved in reducing the proliferation of invading bacteria \cite{proteins1}.
A small oscillation amplitude of $0.2$ nm at $2.2$ kHz was imposed on the tip and the cantilever calibrated.
The calibration curves for this measurement are presented in figure \ref{fig:1}.
Despite a possible contamination of the tip which might induce artifacts in the images, the clear and stable presence of short-range attractive forces between the tip and the sample gives the opportunity to acquire a non-contact image. In the context of biomechanics this is an important instrumental challenge, even if for the moment it does not add anything to our knowledge of the studied system.
The phase difference between excitation and tip oscillations was used as set-point for the acquisition of the topography due to its monotonicity as a function of tip-sample distance. Here the images clearly provide more information than just the topography.
In this case, at variant with the first two examples, the stiffness (Figure 5c) is not linked to the common sample stiffness as such property would imply direct contact with the sample that in this specific situation is absent.
 \begin{figure}[htp]
 \centering
  \includegraphics[width=\linewidth]{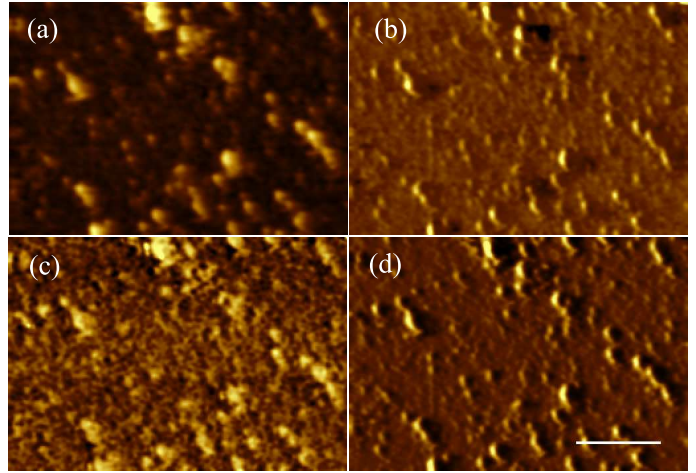} 
 \caption{FFM images of clusters of TBK1 and OPTN complexes deposited on mica 
 in aqueous solution.
 (a) topography, (b) force, (c) stiffness and (d) damping.
 The full color scale corresponds to 24nm, -300pN, -0.04 N/m and 5$\mu$kg/s respectively and
 the scale bar to 1$\mu$m.
 This image was taken in attractive regime using the phase as the feedback signal.}
 \label{fig:4}
 \end{figure}
\section{Discussion}
In all the three cases presented the images are provided with absolute values, based on the experimental cantilever calibration. The error propagation associated to the measurements comes from the cantilever stiffness alone. In conclusion, we have shown that the FFM can provide quantitative images of the mechanical properties of biological samples in liquid media. Depending on the property of interest a particular signal for the feedback loop can be selected. We have shown two configurations: one where the signal to the topography loop is the force and another where the phase is used; other signals such as the amplitude can be used as well. The phase seems to be in general a good candidate because it is often monotonous regardless of the nature of the interaction. The monotonicity is mainly due to the fact that the chosen frequency is far from resonances. In FFM, the high sensitivity is determined by the small cantilever stiffness chosen and not by resonance phenomena. 
\section{Conclusions}
These \emph{proof of principle} experiments underscore the general philosophy of Force Feedback Microscopy and the benefits of the FFM in providing qualitative and quantitative in situ characterization of biological samples. 
Furthermore, the FFM can provide quantitative data on viscoelasticity 
at any given frequency because the choice of working frequency 
is arbitrary and independent of the cantilever resonant mode.
Thus, it is possible to obtain images or approach curves at different 
frequencies to explore the local mechanical impedance of samples. 

\vspace{9pt}
\noindent \textbf{AKNOWLEDGMENTS} \\
Luca Costa acknowledges COST Action TD 1002.
Mario S. Rodrigues acknowledges financial 
support from Fundação para a Ciência e Tecnologia SFRH/BPD/69201/2010.
The authors acknowledge Leide Cavalcanti for the help in preparing the lipid samples.
\hspace{8pt} 

\end{document}